\documentclass[%
reprint,
 amsmath,amssymb
,aps
]{revtex4-2}

\usepackage{amsthm,amsmath,amssymb}
\usepackage{mathrsfs}
\usepackage{graphicx}
\usepackage{hyperref}

\begin{document}

\title{Next-to-leading order corrections to scalar perturbations of Kerr-anti–de Sitter black holes}

\author{Xiang-hao Chu}%
\email{chuxianghao@mail.sdu.edu.cn}
\affiliation{%
 Institute of Frontier and Interdisciplinary Science,
 Key Laboratory of Particle Physics and Particle Irradiation (MOE), Shandong University, Qingdao 266237, China
}%
\author{Yi-qing Chu}%
\email{yiqingchu@mail.sdu.edu.cn}
\affiliation{%
 Institute of Frontier and Interdisciplinary Science,
 Key Laboratory of Particle Physics and Particle Irradiation (MOE), Shandong University, Qingdao 266237, China
}%

\author{Shou-shan Bao}%
\email{ssbao@sdu.edu.cn}
\affiliation{%
 Institute of Frontier and Interdisciplinary Science,
 Key Laboratory of Particle Physics and Particle Irradiation (MOE), Shandong University, Qingdao 266237, China
}%
\author{Hong Zhang}%
\email{hong.zhang@sdu.edu.cn}
\affiliation{%
 Institute of Frontier and Interdisciplinary Science,
 Key Laboratory of Particle Physics and Particle Irradiation (MOE), Shandong University, Qingdao 266237, China
}%
\date{\today}

\begin{abstract}
The small Kerr-anti–de Sitter black hole demonstrates instability due to the superradiance of either a massive or massless scalar field. Previous leading-order approximations of the spectrum are inefficient. In particular, the leading-order real part of the eigenfrequency is insensitive to the spin of the black hole. In this work, we improve the analysis by including the next-to-leading-order contribution. Compared to the numerical calculation, the new spin-dependent real part presents significantly better agreement, and the error in the imaginary part is also reduced to less than 60\% for most black hole spins.
\end{abstract}

\maketitle


\section{Introduction}
Black hole (BH) superradiance is a phenomenon characterized by the amplification of perturbation under specific conditions, resulting from the extraction of rotational energy from BHs. This effect serves as a wave analog to the hypothetical Penrose process~\cite{Penrose:1969pc}, where energy is extracted by a particle orbiting and decaying within the BH’s ergoregion. Superradiance finds application in diverse research domains, encompassing investigations into the stability of rotating BHs ~\cite{Huang:2018qdl,Boskovic:2018lkj,Franzin:2021kvj,Garcia-Saenz:2021uyv,Lin:2021ssw,Guo:2021xao,Arvanitaki:2010sy,Herdeiro:2021znw}. We refer the interested readers to the comprehensive review Brito {\it et al}~\cite{Brito:2015oca} and references therein.

The phenomenon of superradiant scattering in BHs was first identified by Zel’dovich~\cite{Zel'dovich:1971}. A Kerr BH can amplify an incoming scalar wave with frequency $\omega$ under the condition that $\omega < m\Omega_H$~\cite{Bardeen:1972fi,Starobinsky:1973aij}, where $\Omega_H$ represents the angular velocity of the BH outer horizon and $m$ is the azimuthal number. Press and Teukolsky~\cite{Press:1972zz} proposed that if the superradiantly scattered wave is reflected back onto the BH, continuous amplification can occur. This process leads to a reduction in the BH’s angular momentum and rotational energy, a phenomenon commonly referred to as the ``black hole bomb." This concept is frequently used to illustrate the conditions that lead to superradiance instability~\cite{Cardoso:2004nk,Strafuss:2004qc,Hod:2009cp,Press:1972zz,Rosa:2009ei,Witek:2010qc,Dolan:2012yt,Herdeiro:2013pia,Lee:2011ez,Aliev:2014aba,Degollado:2013bha,Hod:2013fvl,Li:2014gfg,Li:2014fna}. To enable this reflection process, a potential well near the BH's event horizon is required to prevent the scattered wave from escaping to infinity. In Kerr spacetime, such a potential well naturally exists for massive bosonic fields.

 The phenomenon of superradiance of BHs in anti–de Sitter (AdS) space has been studied widely \cite{Cardoso:2006wa,Kunduri:2006qa,Kodama:2009rq,Kodama:2007sf,Cardoso:2013pza,Murata:2008xr,Aliev:2008yk,Wang:2014eha,Delice:2015zga,Aliev:2015wla}. Since the boundary of the AdS spacetime behaves like a well, Kerr-AdS BHs exhibit similar superradiance even for massless bosons. Notably, large AdS BHs are generally stable \cite{Hawking:1999dp}, but small rotating \cite{Cardoso:2004hs,Uchikata:2009zz} or charged \cite{Gubser:2008px} AdS BHs in four dimensions display instability due to superradiance.

The computation of superradiance typically involves both numerical and analytical methods. The analytical approach was initially applied to Kerr BHs by Detweiler \cite{Detweiler:1980uk} and later generalized to Kerr-AdS BHs by Cardoso and Dias \cite{Cardoso:2004hs}. This methodology has also been utilized by Li \cite{Li:2012rx}, Delice and Dur$\mathrm{\breve{g}}$ut \cite{Delice:2015zga}, and Aliev \cite{Aliev:2015wla} to Kerr-Newman and higher dimension AdS BHs. These studies clearly demonstrate that small Kerr-AdS BHs are unstable in the presence of ultralight scalar fields. However, the leading-order approximation yields eigenfrequencies that are independent of the BH’s spin, which contradicts numerical results \cite{Uchikata:2009zz}. Mathematically, this discrepancy arises from the use of asymptotic solutions in free AdS spacetime, which are applicable in regions far from the BH and fail to account for the BH's spin.

In this study, we incorporate the BH's spin into the asymptotic solution far from the horizon and generalize the results to the next-to-leading order (NLO) of $1/\ell^2$ and $M\mu$. We also perform numerical calculations using the shooting method. The analytical and numerical results are then compared, showing good agreement.

This paper is structured as follows: In Sec.~\ref{sec:kerrads}, we introduce the fundamental equations and discuss the superradiance condition of BHs. In Sec.~\ref{sec:lo}, we provide a brief review of the leading-order calculation. In Sec.~\ref{sec:nlo}, we present the results with the NLO correction. In Sec.~\ref{sec:num}, we compare the NLO result with numerical findings. Finally, we summarize our conclusions in Sec.~\ref{sec:sum}. Throughout this article, we adopt natural units with $G=\hbar=c=1$.

\section{Scalar field in Kerr-AdS spacetime}\label{sec:kerrads}
\subsection{Kerr-AdS BHs}
The line element of Kerr-AdS BH spacetime in the Boyer-Lindquist-type coordinates is expressed in the form~\cite{Carter:1968rr,Carter:1968ks,Hawking:1998kw},
\begin{equation}
\begin{aligned}
		ds^{2}=&-\frac{\Delta_{r}}{\Sigma}\left(d t-a\frac{\sin^{2}\theta}{\Xi}d\phi\right)^{2}+\frac{\Sigma}{\Delta_{r}}d r^{2}+\frac{\Sigma}{\Delta_{\theta}}d\theta^{2} \\
		&+{\frac{\Delta_{\theta}\sin^{2}\theta}{\Sigma}}\Biggl(a d t-{\frac{r^{2}+a^{2}}{\Xi}}d\phi\Biggr)^{2},
\end{aligned}
\end{equation}
where
\begin{equation}
\begin{aligned}
	\Sigma&= r^{2}+a^{2}\mathrm{cos}^{2}\theta,\qquad \Delta_{\theta}=1-\frac{a^{2}}{\ell^{2}}\cos^{2}\theta,\\
	\Delta_{r}&=\left(r^{2}+a^{2}\right)\left(1+{\frac{r^{2}}{\ell^{2}}}\right)-2 M r 
	,\quad \Xi=1-\frac{a^{2}}{\ell^{2}}.
\end{aligned}
\end{equation}
The parameter $\ell$ denotes the curvature length, which is related to the negative cosmological constant in AdS space, given as $\Lambda=-3/\ell^2$. 
The $M$ and $a$ are the parameters related to the mass and spin of the BH. The physical mass and angular momentum are expressed as $M/\Xi^2$ and $Ma/\Xi^2$, respectively~\cite{Gibbons:2004ai}. 

The horizons are located at the two real roots of the equation $\Delta_r=0$, denoted as $r_+$ and $r_-$. To ensure the existence of these real roots, a constraint on $M$ is imposed as follows:
\begin{align}
		M>M_c=&\frac{\ell}{3\sqrt{6}}\left\{2\left(1+\frac{a^{2}}{\ell^{2}}\right)+\sqrt{\left(1+\frac{a^{2}}{\ell^{2}}\right)^{2}+\frac{12a^{2}}{\ell^{2}}}\right\}\nonumber \\
		&\times\sqrt{\sqrt{(1+\frac{a^2}{\ell^2})^2+\frac{12a^2}{\ell^2}}-(1+\frac{a^2}{\ell^2})}.\label{eq:horizon}
\end{align}
This constraint places an upper limit on the spin $a$ of the Kerr-AdS BH. In the limit of $\ell\to\infty$ or $\Lambda\to 0$, the constraint reduces to that of a Kerr BH, i.e. $a<M$. Additionally, to avoid singularities in $\Delta_\theta$, the BH spin is constrained to be
\begin{align}
 a<\ell.\label{eq:singulary}
\end{align}
For small BHs with $M\ll \ell$, we can expand the horizon condition in terms of $1/\ell$ and obtain the upper limit on the Kerr-AdS BH spin,
\begin{align}
 a<a_\mathrm{max}\sim M\left(1-\frac{M^2}{\ell^2}\right)+\mathcal{O}(M^3/\ell^3), \label{eq:horizonapp}
\end{align}
which is more stringent than that in Eq.~\eqref{eq:singulary} for small BH $M<\ell$.
 
The full expressions for the horizons $r_+$ and $r_-$ are quite complicated and we do not present them here. In addition to the two real roots, the equation $\Delta_r=0$ has two complex roots which are conjugate to each other. Consequently, the $\Delta_r$ can be factored as,
\begin{align}
\Delta_r&=(r-r_+)(r-r_-)\left(\frac{r^2}{\ell^2}+\frac{r_++r_-}{\ell^2} r+\frac{a^2}{r_+r_-}\right).
\end{align}
The two complex roots $\hat{r}_{\pm}$ can be expressed with $r_\pm$,
\begin{align}
\hat{r}_{\pm}=-(\hat{\alpha}\pm i \hat{\beta}),
\end{align}
where
\begin{align}
&\hat{\alpha}=\frac{r_+ + r_-}{2},\quad \hat{\beta}^2=\frac{a^2 \ell^2 }{r_+ r_-}-\hat{\alpha}^2.
\end{align}
On the other hand, the spin $a$ and mass $M$ can also be expressed as functions of $r_\pm$,
\begin{align}
a^2 &={r_+ r_-}\times \frac{1+(r_+^2+r_-^2+r_+ r_-)/\ell^2}{1- r_+ r_-/\ell^2},\\
M&=\frac{r_+ +r_-}{2} \times \frac{(1+r_+^2/\ell^2)(1+r_-^2/\ell^2)}{1-r_+r_-/\ell^2}.
\end{align}
These expressions allow us to use $r_\pm$ instead of $a$ and $M$ as the parameters of the BH, which is more convenient when analyzing the asymptotic behavior near the horizon without expansion with $1/\ell$.

\begin{figure}[!tb]
	\centering
	\includegraphics[scale=0.4]{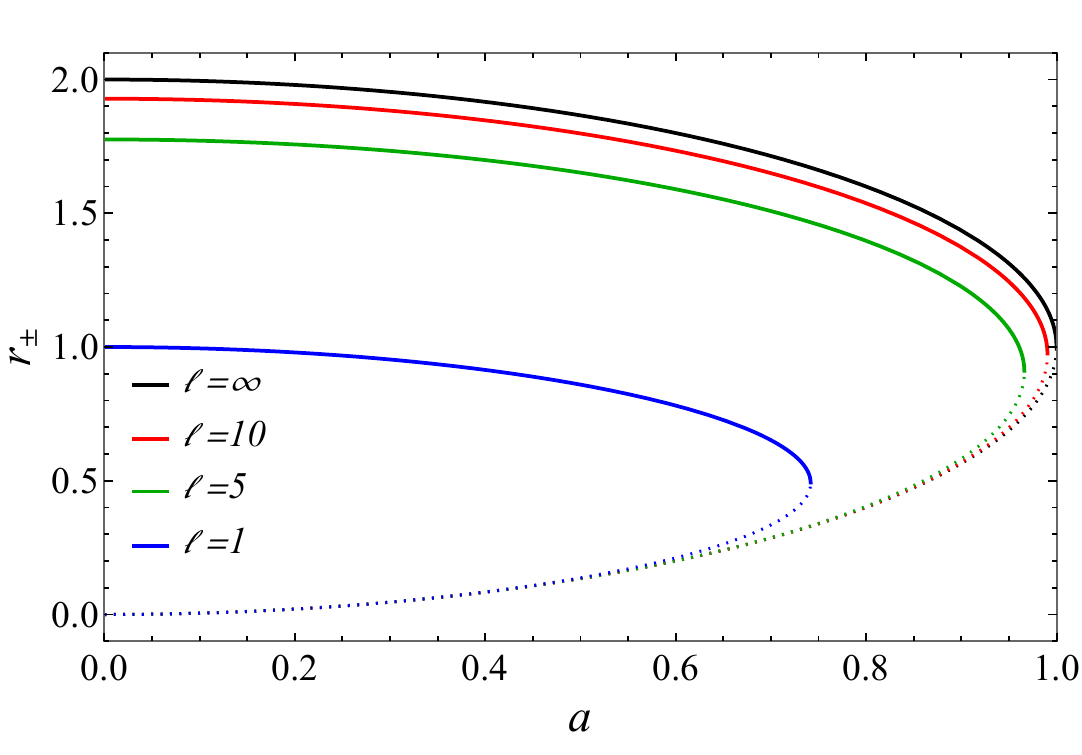}
	\caption{\label{fr1} The outer horizon $r_+$ (solid curves) and the inner horizon $r_-$ (dotted curves) of the Kerr-AdS BH for different $\ell$. Different colors represent different $\ell$ (see legend). The curves with $\ell=\infty$ are the horizons of a Kerr BH.}
\end{figure}

Figure.~\ref{fr1} illustrates $r_+$ and $r_-$ as functions of $a$ with different values of $\ell$. We can see that the  $r_+$ of a Kerr-AdS BH is always less than that of the Kerr BH. For a fixed value of $\ell$, the two horizons $r_+$ and $r_-$ converge at the maximum value of $a$ given by  Eq.~\eqref{eq:horizonapp}. When $\ell$ approaches infinity, a Kerr-AdS BH reduces to a Kerr BH. The difference between the horizons of Kerr-AdS and Kerr BH can be expressed as,
\begin{align}
r_{\pm}-r_{\pm}^K=-\frac{M(\sqrt{M^2-a^2}\pm M)^3}{\ell^2\sqrt{M^2-a^2}}+\mathcal{O}(1/\ell^4),\label{eq:rpmdiff}
\end{align}
where the $r_{\pm}^K$ denote the Kerr BH horizons. The difference decreases as $\ell^{-2}$, inspiring previous studies to approximate a Kerr-AdS BH by a Kerr BH near the horizon. Nonetheless, the difference only decreases as $\ell^{-1}$ when the BH spin approaches its limiting value given in Eq.~\eqref{eq:horizonapp}. In this case, the effect of $\ell$ near the horizon cannot be simply neglected. We discuss the effect in Sec.~\ref{sec:lo}.

\subsection{Klein-Gordon equation}\label{sec:II}

The dynamics of a free real scalar field with mass $\mu$ in the Kerr-AdS background is described by the Klein-Gordon equation,
\begin{equation}
	\left( \nabla^{2}-\mu^2\right) \Phi=0. \label{eq:kgeq}
\end{equation}
The wave function $\Phi(t,r,\theta,\phi)$ can be factorized as~\cite{Carter:1968ks,Teukolsky:1972my}
\begin{equation}
\Phi=\displaystyle\sum_{lm}\int d\omega\left[ e^{-i\omega t}e^{im\phi}R_{lm}(r)S_{lm}(\theta)+c.c.\right],
\end{equation}
where $l$ and $m$ are the azimuthal and magnetic numbers, respectively. The frequency $\omega$ can be either real or complex. The sign of the imaginary part of $\omega$ determines whether the solution is decaying (Im$(\omega)<0$) or growing (Im$(\omega)>0$) in time. Inserting this expression into Eq.~\eqref{eq:kgeq}, one obtains the radial and angular equations,
\begin{equation}
\begin{aligned}
	\Delta_r\frac{d}{dr}\left(\Delta_r\frac{dR}{dr}\right)+ \Big[ \left( \omega\left( r^2+a^2\right) -ma\Xi\right) ^2&\\
	-\left( \mu^2r^2+\Lambda_{lm} \right) \Delta_r \Big]R_{lm}&=0, \label{eq:radial}
\end{aligned}
\end{equation}
\begin{equation}
\begin{aligned}
	\frac{\Delta_{\theta}}{\sin\theta}\frac{d}{d\theta}\left(\Delta_{\theta}\sin\theta\frac{dS}{d\theta}\right) -\Bigg[ \left(\omega a\sin\theta-\frac{m\Xi}{\sin\theta}\right)^2&\\
	+\mu^2 a^2\Delta_{\theta}\cos^2\theta-\Lambda_{lm}\Delta_{\theta}\Bigg] S_{lm}&=0, \label{eq:angular}
\end{aligned}
\end{equation}
where the separation constant $\Lambda_{lm}$ can be solved numerically with a continued fraction equation~\cite{Uchikata:2009zz}. For nonrotating BHs, the angular wave function $S_{lm}$ reduces to spherical harmonics and $\Lambda_{lm}=l(l+1)$. In this work, we
consider the case with both $a^2(\omega^2-\mu^2)$ and $a/\ell$ are small; then, the $\Lambda_{lm}$ could be expanded as~\cite{Suzuki:1998vy,Cho:2009wf,Berti:2005gp,Seidel:1988ue}
\begin{equation}
\begin{aligned}
\Lambda_{lm}=& l(l+1)+a^2\mu^2-2am\omega-\frac{14a^2}{5\ell^2}\\
&+\frac{4a^2(\omega^2-\mu^2)}{5}+{\cal O}\left(a^4/\ell^4,a^4\left(\omega^2-\mu^2\right)^2\right). \label{eq:lambdalm}
\end{aligned}
\end{equation}

\subsection{Superradiance condition}
To discuss the asymptotic behavior of the scalar field, it is useful to introduce the tortoise coordinate $r_*$ which is defined as
\begin{align}
\frac{dr_*}{dr}=\frac{r^2+a^2}{\Delta_r}.
\end{align}
The radial equation can be expressed as
\begin{align}
\frac{d^2 R_*}{dr_*}&+\left[\left(\omega-\frac{am\Xi}{r^2+a^2}\right)^2\right.\nonumber\\
&\left.-\frac{\Delta_r(\mu^2 r^2+\Lambda_{lm})}{(r^2+a^2)^2}-G^2-\frac{dG}{dr_*}\right]R_*=0,\label{eq:schrodinger}
\end{align}
with
\begin{align}
R_*=\sqrt{r^2+a^2} R(r) \quad \text{and}\quad G=\frac{\Delta_r r}{(r^2+a^2)^2}.
\end{align}
This is equivalent to the one-dimensional scattering problem. Near the event horizon $r\to r_+$, we have $\Delta_r\to 0$ and the radial equation becomes
\begin{align}
\frac{d R_*}{dr_*^{2}}+\left(\omega-m \Omega_+\right)^2 R_*=0,
\end{align} 
where $\Omega_+$ is the angular velocity of the outer horizon measured relative to a frame rotating at infinity,
\begin{align}
\Omega_+=\frac{a}{r_+^2+a^2}\Xi.
\end{align}
The wave function $R_*$ at $r\to r_+$ behaves as
\begin{align}
R_*(r\to r_+)\sim e^{\pm i (\omega-m \Omega_+) r_*},\label{eq:tranfer}
\end{align}
where the $+$ sign represents an unphysical outgoing wave at the horizon. Thus, we have the asymptotic solution near the horizon as
\begin{align}
R_*(r\to r_+)=B e^{- i (\omega-m \Omega_+) r_*},
\end{align}
where $B$ is the normalization factor to be determined. 

For Kerr BHs, the superradiance condition can be obtained by comparing the amplitudes of the incident wave and reflected wave at $r\to\infty$. This method cannot be applied to AdS spacetime directly, which is not asymptotic flat at $r\to \infty$.  Nevertheless, in limit of $\ell\gg r_+$, the spacetime in the region $r_+\ll r\ll \ell$ is asymptotic flat, and the radial equation is
\begin{align}
&\frac{d R_*}{dr_*^{2}}+\left(\omega^2-\tilde{\mu}^2\right) R_*=0, 
\end{align}
with
\begin{align}
&\tilde{\mu}^2=\mu^2+\frac{\Lambda_{lm}+2}{\ell^2}.\label{eq:mus}
\end{align} 
The asymptotic solution can be composited with two plane waves,
\begin{align}
R_*= e^{-i k r_*}+Ae ^{i k r_*},
\end{align}
with $k=\sqrt{\omega^2-\tilde{\mu^2}}$.
This solution describes the incident wave with unit amplitude and reflected wave with the amplitude $|A|$, while Eq.~\eqref{eq:tranfer} is the transmitted wave. From the conservation of the Wronskian of the solutions, one could obtain
\begin{align}
1-|A|^2&=\frac{\omega-m\Omega_+}{\sqrt{\omega^2-\tilde{\mu}^2}} |B|^2, 
\end{align}
The occurrence of the superradiance $|A|>1$ requests
\begin{align}
\omega<\frac{am\Xi}{r_+^2+a^2}.\label{eq:superradiance}
\end{align}
Using the fact that $\omega\sim\tilde{\mu}$, this superradiance condition reduces to the Kerr BH situation in the limit of $\ell\to \infty$. 

\begin{figure}[!tb]
	\centering
	\includegraphics[scale=0.4]{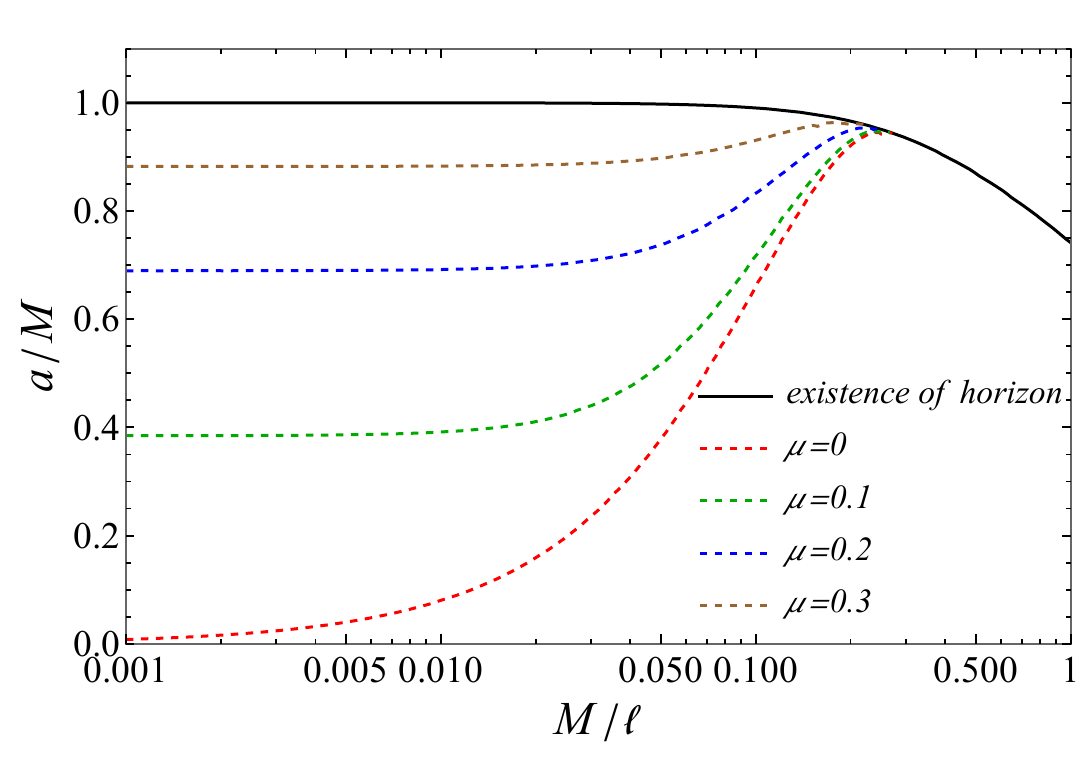}
	\caption{\label{f1}Superradiance conditions for different values of $\mu$ with $n=0$ and $l=m=1$. The normalization is chosen as $M=1$. The solid curve is the upper limit for the existence of the horizons, while the dashed curves represent the superradiance conditions for different $\mu$. Given $\mu$, the parameter space between the corresponding dashed curve and the solid curve is allowed for superradiance.}
\end{figure}

By assuming  $\omega=\tilde{\mu}$, we demonstrate the superradiance conditions for different values of $\mu$ in Fig.~\ref{f1}. The black solid curve is the horizon condition given in  Eq.~\eqref{eq:horizon}, and the dashed curves are the superradiance bounds given in  Eq.~\eqref{eq:superradiance} with different $\mu$. The region between the solid and dashed curves is the parameter space where superradiance is plausible. For larger scalar mass $\mu$, higher BH spin $a$ is required for superradiance. When $\ell\to \infty$, which is the Kerr BH case, one could obtain the upper value of $\max(\mu)=1/2$. Meanwhile, one could also observe the dashed curves merge with the solid one at some value of $M/\ell$. With BH mass $M$ fixed, it provides a lower limit on $\ell$. The value can be estimated from Eq.~\eqref{eq:superradiance} and  Eq.~\eqref{eq:mus}. With $\mu=0$, $M=1$ and $l=m=1$, one obtains
\begin{align}
\ell>\sqrt{2+\Lambda_{lm}}\left(\frac{r_+^2+a^2}{am\Xi}\right)\sim 4.
\end{align}

\section{Review of the LO Solution}\label{sec:lo}
In this section, we provide a concise overview of leading-order (LO) approximation following Ref.~\cite{Cardoso:2004hs} and incorporate the contribution with nonzero $\mu$.  In the subsequent section, we generalize the calculation by including the higher-order corrections. 

\subsection{Near-region solution}

In the region near the horizon $r_+$, we introduce a new variable $z=\left( r-r_+\right) /\left( r_+-r_-\right) $; the radial equation  can then be rewritten in terms of $z$, 
\begin{equation}
	z\left( z+1\right) \frac{d}{dz}\left[ z\left( z+1\right) \frac{dR}{dz}\right] +V(z)R=0, \label{eq:near}
\end{equation}
where
\begin{align}
	V(z)&=\left[\frac{a^2+r_+^2}{r_+-r_-}\left( \omega-m\Omega_+\right) \right]^2  \nonumber\\
	&-z\left[\Lambda_{lm}+\mu^2 r_+^2-\frac{4r_+\omega (r_+^2+a^2) (\omega-m\Omega_+)}{r_+-r_-}\right]\nonumber\\
	&-z^2\Big[\Lambda_{lm}+3\mu^2 r_+^2-2 \mu^2r_+r_- -4\omega^2r_+^2 \nonumber\\
	&\quad -2\omega\left(r_+^2+a^2\right)\left(\omega-m\Omega_+\right)\Big]+\mathcal{O}\left( z^3\right).\label{eq:near_v}
\end{align}
At the LO of $M\mu$, we only consider the leading terms $\Lambda_{lm}\sim l(l+1)$ in the coefficients of $z$ and $z^2$. Then the potential can be reduced to,
\begin{align}
V(z)={p}^2-l'(l'+1)z(z+1), \label{eq:near_LO_V}
\end{align}
where
\begin{align}
p=\frac{a^2+r_+^2}{r_+-r_-}\left( \omega-m\Omega_+\right). \label{eq:p}
\end{align}
The $l^\prime$ is defined as $l^\prime=l+\epsilon$, where $\epsilon$ is a regulator which is taken to be zero at the end at LO.  
The solution with the ingoing boundary condition is given as
\begin{align}
	R(r)\sim&\left( \frac{r-r_+}{r-r_-}\right) ^{-ip}\nonumber\\
	&\times {}_2F_1\left( -l',l'+1;1-2ip;\frac{-r+r_+}{r_+-r_-}\right). 
\end{align}
The large-$r$ behavior of this solution is
	\begin{align}
	R\sim&\frac{\left( r_+-r_-\right) ^{-l'}\Gamma\left( 2l'+1\right) }{\Gamma\left(l'+1\right)\Gamma\left( l'+1-2ip\right) }r^{l'}\nonumber\\
	&+\frac{\left( r_+-r_-\right) ^{l'+1}\Gamma\left(-2l'-1\right) }{\Gamma\left(-l'\right) \Gamma\left( -l'-2ip\right)}r^{-l'-1}. \label{eq:near_sol}
	\end{align}
This expression is well-defined with the regulator $\epsilon$. One could take the limit of $\epsilon\to 0$ after calculation of the irregular piece,
\begin{align}
\lim_{\epsilon\to0}
\frac{\Gamma(-2l^\prime-1)}{\Gamma(-l^\prime)}=\frac{1}{2}\times\frac{(-1)^{l+1} l!}{(2l+1)!}.
\end{align}
Finally by taking $\epsilon\to 0$ limit, Eq.~\eqref{eq:near_sol} is simplified as
	\begin{align}
	R\sim&\frac{r^{l} \left( r_+-r_-\right) ^{-l} (2l)! }{ l! \Gamma\left( l+1-2ip\right) }+\frac{r^{-l-1} \left( r_--r_+\right) ^{l+1} l! }{2(2l+1)!\Gamma\left( -l'-2ip\right)}. \label{eq:near_LO}
	\end{align}

\subsection{Far-region solution}
In the far region, $r-r_+\gg M$, the radial wave equation reduces to the wave equation of a scalar field in a pure AdS background at the LO of $1/\ell$,
\begin{equation}
\begin{aligned}
		\left( 1+\frac{r^2}{\ell^2}\right) \partial_{r}^{2}R(r)+{{2r\left(\frac{2}{\ell^{2}}+\frac{1}{r^2}\right)\partial_{r}R(r)}}&\\
		+\left( \frac{\omega^2}{1+r^2/\ell^2}-\frac{l(l+1)}{r^2}-\mu^{2}\right)R(r)&=0. \label{eq:farLO}
\end{aligned}
\end{equation}
It can be rewritten in a standard hypergeometric form by choosing a new variable $x=1+r^2/\ell^2$,
\begin{align}\label{eq:far_eq_near}
&x(1-x)\partial_{x}^{2}R+\frac{2-5x}{2}\partial_{x}R\nonumber\\
&\quad -\left[\frac{\omega^{2}\ell^{2}}{4x}+\frac{l(l+1)}{4(1-x)}-\frac{\mu^2\ell^2}{4}\right]R=0.
\end{align}
The solution of the equation at $x\to \infty$ is a linear combination of two hypergeometric functions,
	\begin{equation}\label{eq:far_sol_near}
	\begin{aligned}
		R(x)&=\left(x-1\right) ^{\frac{l}{2}} x^{\frac{1}{2}\omega\ell}\times\\
		& \left[ C_1 x^{-\alpha }{}_2F_1\left( \alpha,\alpha-\gamma+1,\alpha-\beta+1,x^{-1}\right)\right. \\
		&\left.+C_2 x^{-\beta}{}_2F_1\left( \beta,\beta-\gamma+1,\beta-\alpha+1,x^{-1}\right)\right] ,
	\end{aligned}
\end{equation}
with the parameters given by
\begin{subequations}\label{eq:abg}
\begin{align}
		\alpha&=\frac{l}{2}+\frac{3}{4}+\frac{1}{4}\sqrt{9+4\mu^{2}\ell^{2}}+\frac{\omega \ell}{2}, \\
		\beta&=\frac{l}{2}+\frac{3}{4}-\frac{1}{4}\sqrt{9+4\mu^{2}\ell^{2}}+\frac{\omega \ell}{2}, \\
		\gamma&=1+\omega\ell. 
\end{align}
\end{subequations}
In the massless limit $\mu=0$, the expressions of $\alpha$, $\beta$, and $\gamma$ reduce to those in Ref.~\cite{Cardoso:2004hs}. 
In this work, we consider the bound state, which approaches to be zero at infinity, i.e.,  $R(r\to\infty)\to 0$. As a result, there should be $C_2 =0$. The scenarios with other boundary conditions~\cite{Ishibashi:2004wx} are beyond this work. 
It is worth noting that when the $\alpha-\beta$ is an integer, the general solution to Eq.~\eqref{eq:far_eq_near} is more complicated. However, given the boundary condition under consideration, the solution in Eq.~\eqref{eq:far_sol_near} with $C_2=0$ remains valid. 

Taking $C_2=0$ and using the property of the hypergeometric function, one obtains the small-$r$ behavior of the solution in Eq.~\eqref{eq:far_sol_near}, 
 \begin{equation}
 	\begin{aligned}
 	R\sim &\frac{\Gamma\left( \gamma-\alpha-\beta\right) }{\Gamma\left( 1-\beta\right) \Gamma\left( \gamma-\beta\right) }\left(\frac{r}{\ell}\right)^{l}\\
 	&+\frac{\Gamma\left( \alpha+\beta-\gamma\right) }{\Gamma\left( \alpha\right) \Gamma\left( \alpha-\gamma+1\right) }\left(\frac{r}{\ell}\right)^{-l-1}. \label{eq:far_LO}
 	\end{aligned}
 \end{equation}
To have a regular behavior of the solution at the limit of $r\to 0$, the coefficient of $r^{-l-1}$ in Eq.~\eqref{eq:far_LO} must be severely suppressed. It means $\alpha-\gamma+1$ is in the neighborhood of zero or some negative integer,
\begin{equation}
\alpha-\gamma+1=-n-\delta, \label{eq:deltabeta}
\end{equation}
where $\vert\delta\vert \ll 1$ and integer $n\geq0$. With $\alpha$ and $\gamma$ given in Eq.~\eqref{eq:abg}, one could obtain \cite{Cardoso:2004hs,Berti:2005gp,Breitenlohner:1982bm,Ishibashi:2004wx},
\begin{align}\label{eq:omega_LO_delta}
\omega=\frac{4n+2l+3+\sqrt{9+4\mu^2\ell^2}+4\delta}{2\ell}.
\end{align}
Once $\delta$ is known, we can obtain the eigenfrequency $\omega$ from this expression. The value of $\delta$ can be obtained by matching the two solutions in Eqs.~\eqref{eq:near_LO} and  \eqref{eq:far_LO}, which is explained below. 


\subsection{Matching}
In the case of $r_+\omega\sim r_+\mu\ll1$ and $a^2/\ell^2\ll1$, there is a region in which both Eqs.~\eqref{eq:near_LO} and  \eqref{eq:far_LO} are valid. By matching the two expressions in this region, one arrives at
\begin{align}\label{eq:delta_LO}
		\delta&=\frac{(-ip)}{ \sqrt{\pi} \ell^{2l+1}}\frac{2^{l-n} (r_+-r_-)^{2l+1} (l!)^{2} (2n+2l+1)!!}{n!(2l+1)!(2l)!(2l+1)!!(2l-1)!!} \nonumber\\
		&\times\frac{\Gamma(n+l+3/2+\sqrt{9/4+\mu^{2}\ell^{2}})}{\Gamma(n+1+\sqrt{9/4+\mu^{2}\ell^{2}})}\left[\prod_{k=1}^{l}\left( k^{2}+4p^{2}\right) \right]. 
\end{align}
This expression is consistent with Ref.~\cite{Uchikata:2009zz} in the case of $\mu=0$ and Refs.~\cite{Delice:2015zga,Aliev:2015wla} when the spacetime dimension is adjusted to $4$, but differs by a factor $1/2$ with Ref.~\cite{Cardoso:2004hs} in the case of $\mu=0$ and  Ref.~\cite{Li:2012rx} when the BH charge is taken to $0$. The  overall factor comes from a mistake in treating the Gamma functions with negative integer arguments, which is explored for Kerr BHs in Ref.~\cite{Bao:2022hew}.

One could then calculate the eigenfrequency by inserting Eq.~\eqref{eq:delta_LO} into Eq.~\eqref{eq:omega_LO_delta}. Note $\delta$ also depends on $\omega$. Since $\delta$ is small, it could be treated perturbatively. Mathematically, one has
\begin{align}
\omega_0&=\frac{4n+2l+3+\sqrt{9+4\mu^2\ell^2}}{2\ell},\\
\omega_\text{LO}&=\omega_0+\frac{2}{\ell}\delta(\omega_0),\label{eq:omega_lo}
\end{align}
where the subscript ``LO" represents the LO of the imaginary part of $\omega$, and $\delta(\omega_0)$ means taking $\omega=\omega_0$ in the expression of $\delta$ in Eq.~\eqref{eq:delta_LO}.

\begin{figure}[!tb]
	\centering
	\includegraphics[width=0.4\textwidth]{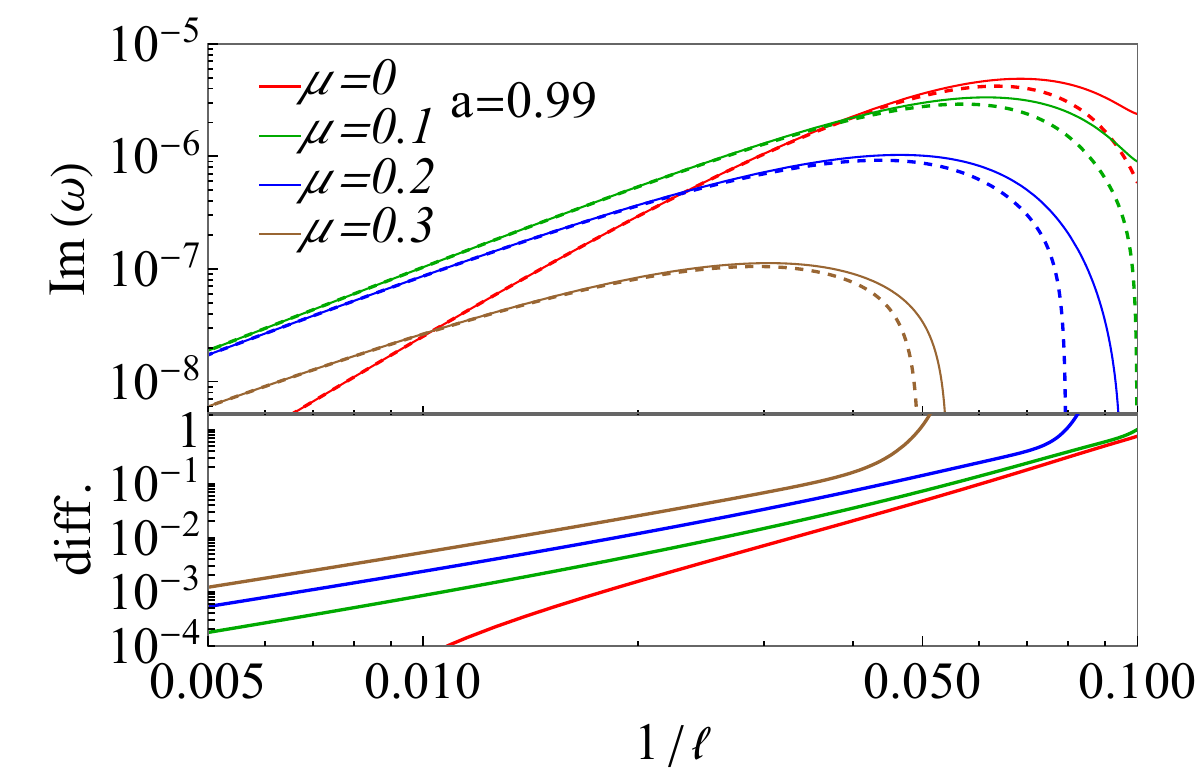} \includegraphics[width=0.4\textwidth]{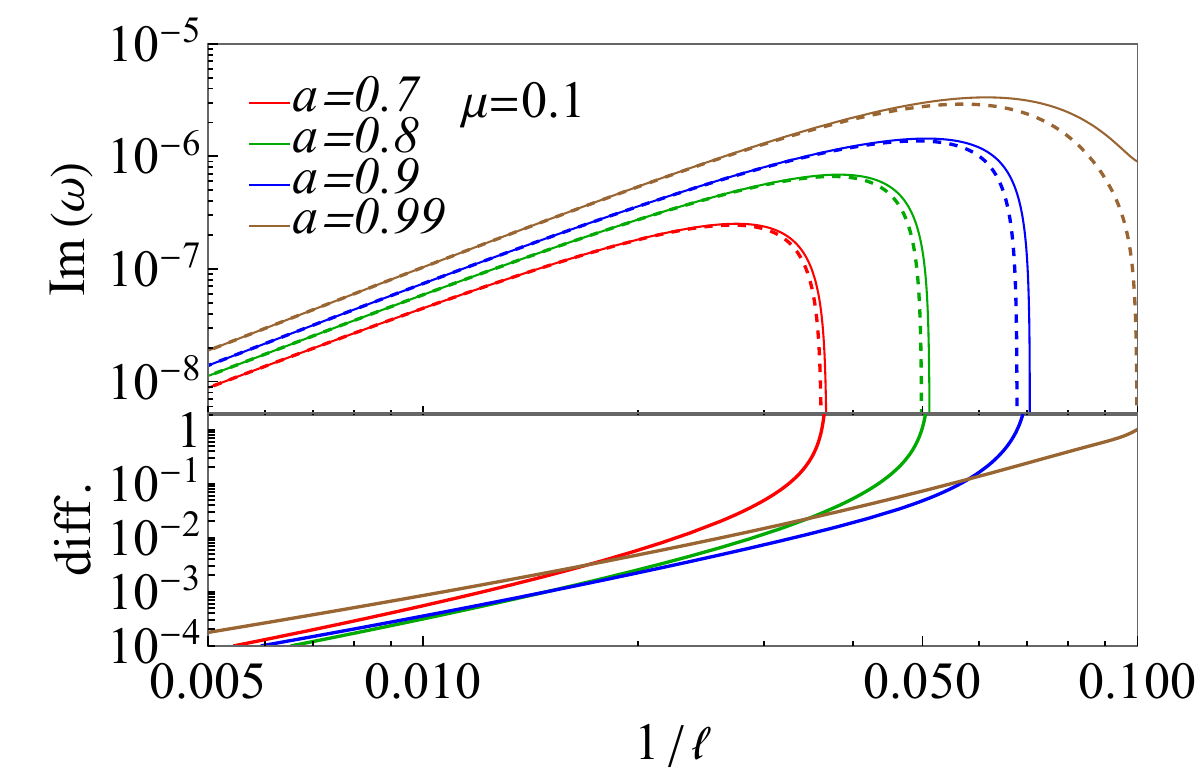}
	\caption{The LO results of the $\mathrm{Im}(\omega)$ (solid curves) for different $\ell$ with $a=0.99$, $M=1$, $n=0$ and $l=m=1$. For comparison, the LO results with $r_\pm$ replaced by $r_\pm^K$ are presented in dashed curves and are referred to as LO$_K$. Their relative difference is defined as $1-\mathrm{Im}(\omega_\mathrm{LO_{K}})/\mathrm{Im}(\omega_\mathrm{LO})$.}\label{fig:omegaLO}
\end{figure}

Next, we discuss some properties of this LO result. First, one may expect the Kerr-AdS BH reduces to the Kerr BH in the $\ell\to\infty$ limit. Nonetheless, the expression of $\delta$ in Eq.~\eqref{eq:delta_LO} approaches zero in this limit, leading to a real eigenfrequency, which contradicts the case of the Kerr BH. The answer is clear when we revisit the power counting rule leading to Eq.~\eqref{eq:delta_LO}. To ignore the $\mathcal{O}(z^3)$ terms in Eq.~\eqref{eq:near_v}, we have assumed $r\ll \mu^{-2}$. For simplicity , we take $M=1$ here. More details of the power counting can be found in Refs.~\cite{Bao:2022hew,Bao:2023xna}. On the other hand, Eq.~\eqref{eq:farLO} requires $r\gg \ell \sqrt{\mu}$. For these two equations having an overlapping range of $r$, we have implicitly assumed $\ell <\mu^{-5/2}$. Since $\delta$ in Eq.~\eqref{eq:delta_LO} is valid only in this parameter space, it is not surprising that it does not reduce to the Kerr case with $\ell$ approaching infinity.

Second, Eq.~\eqref{eq:delta_LO} depends on $r_+-r_-$, which receives modifications of $\ell$.
The difference between the horizons with the Kerr horizons is shown in Fig.~\ref{fr1} and written in an expanded form in Eq.~\eqref{eq:rpmdiff}. In Refs.~\cite{Cardoso:2004hs,Li:2012rx}, the Kerr horizons are used in Eq.~\eqref{eq:delta_LO} to simplify the calculation. In Fig.~\ref{fig:omegaLO}, we evaluate this simplification numerically and calculate the percentage error. As expected from Eq.~\eqref{eq:rpmdiff}, the difference is important with a small value of $\ell$ or when the BH spin is close to its maximum value. For the maximum value of $\text{Im}(\omega)$ and $l=1$, replacing the Kerr-AdS horizons with the Kerr ones leads to an error at the order of $10\%$.
The error is larger with the larger value of $l$.

Finally, by inserting $\omega_0$ in Eq.~\eqref{eq:omega_lo} into the superradiance condition in Eq.~\eqref{eq:superradiance}, we could obtain a critical BH spin below which the superradiance could not happen,
\begin{align}
a^{nlm}_c
\simeq\frac{ r_+^2\omega_0^{nlm} }{m} \left(1+ \frac{r_+^2\omega^{nlm}_0 }{m^2}\right). \label{eq:a_critical}
\end{align}
The critical spin of superradiance plays an important role in the analysis of the evolution of the BH and the scalar field~\cite{Guo:2022mpr}.

\section{NLO correction}\label{sec:nlo}
\begin{figure*}[!tb]
	\centering
	\includegraphics[width=0.3\textwidth]{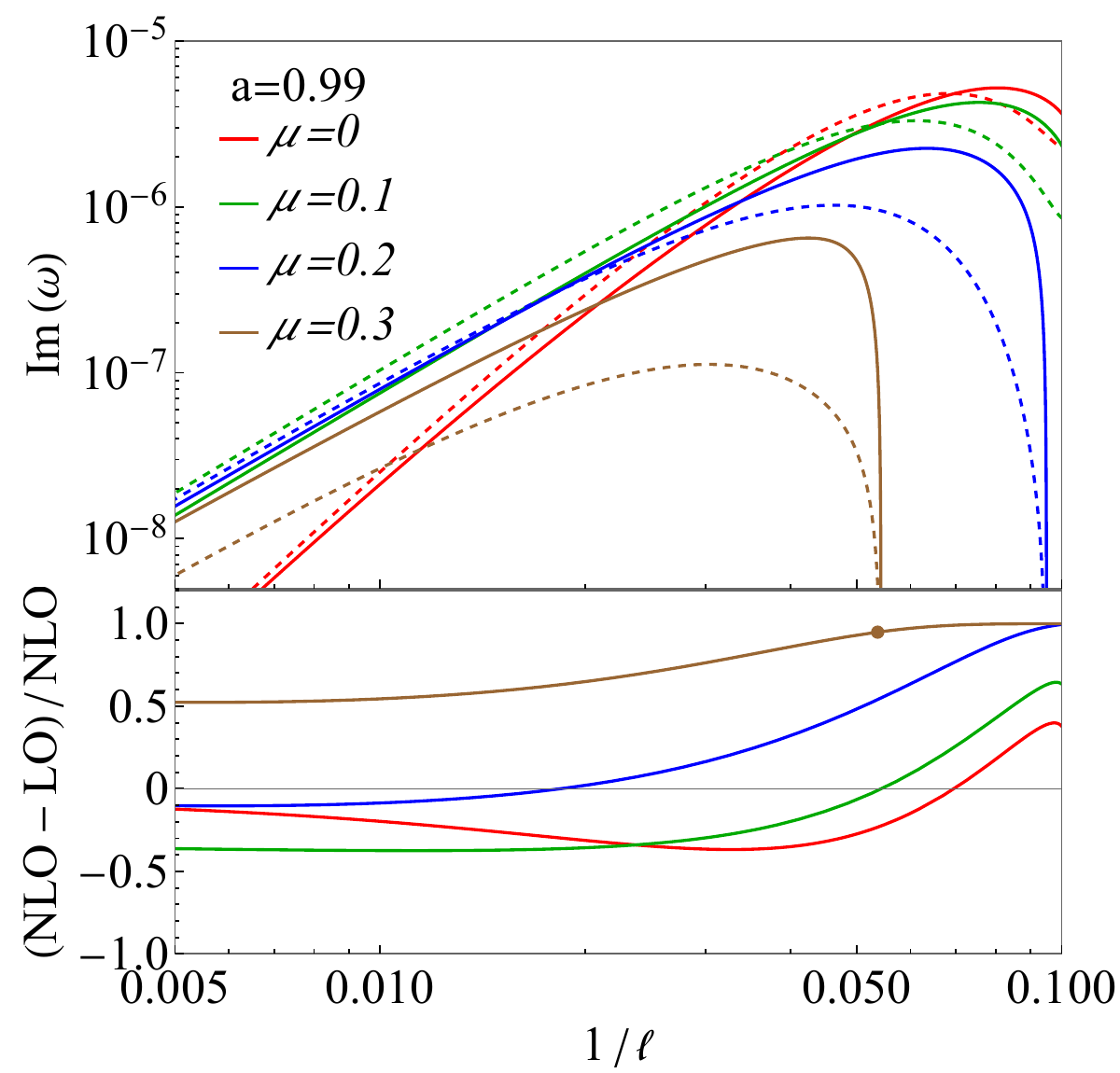}
	\includegraphics[width=0.3\textwidth]{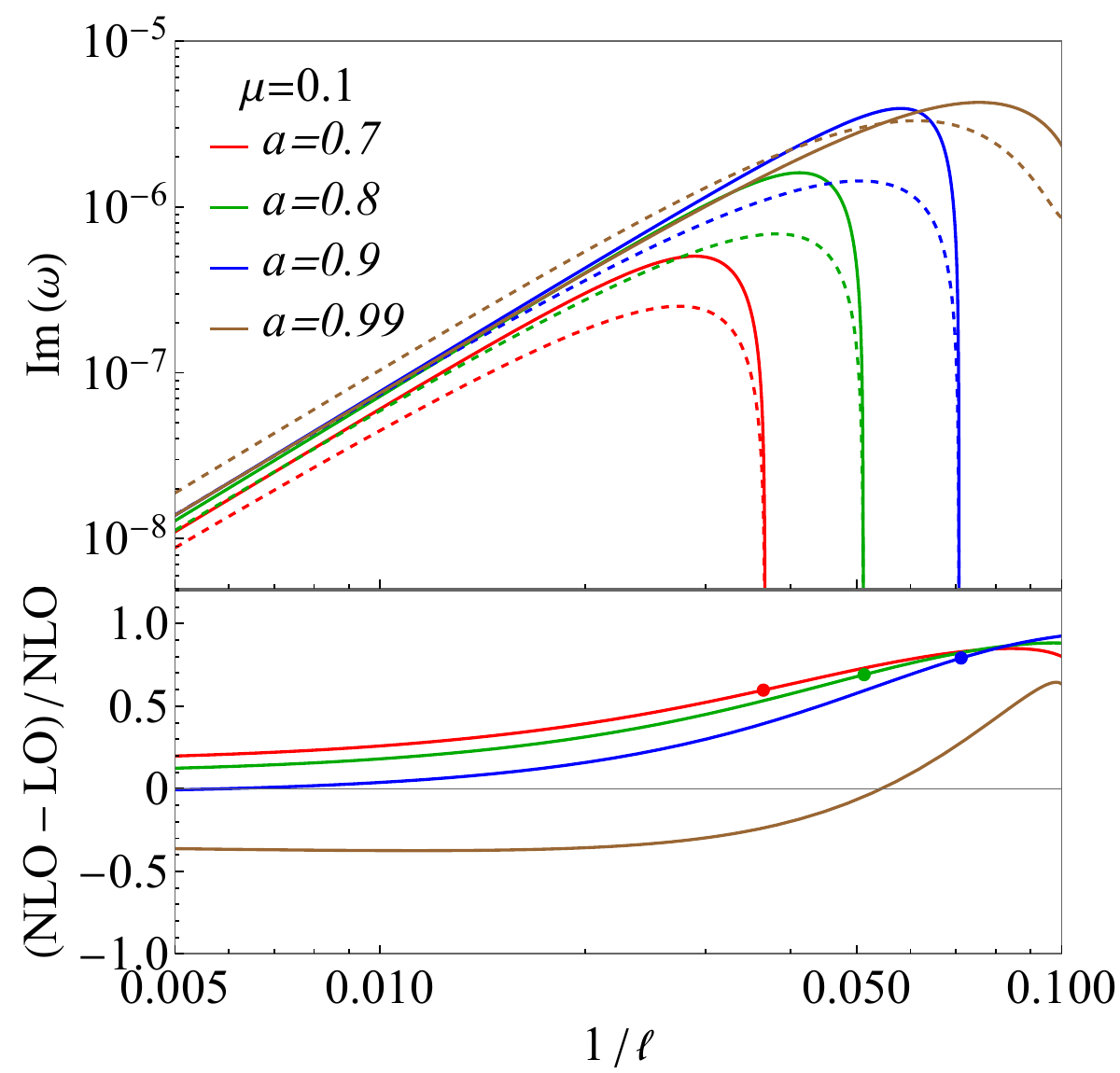}
	\includegraphics[width=0.3\textwidth]{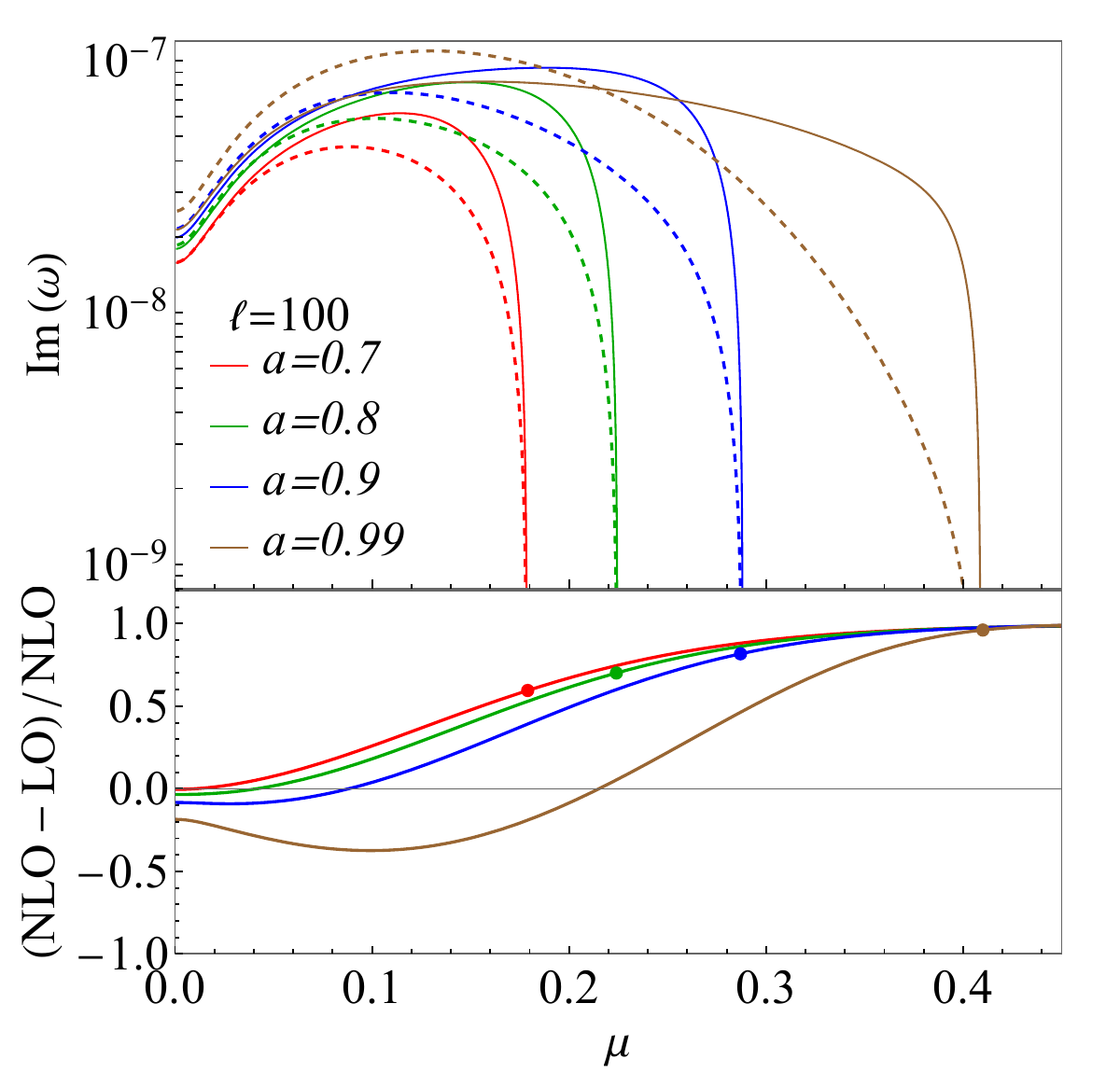}
	\caption{\label{fig:lo_nlo} The LO and NLO results of the $\mathrm{Im}(\omega)$ and their relative difference as a function of $\ell$ and $\mu$ with $a=0.99$, $n=0$ and $l=m=1$. The normalization is chosen as $M=1$. In the top panels, the solid and dashed curves denote the LO and NLO results, respectively.}
\end{figure*}

In this section, we revisit the calculation above to obtain the NLO correction. Unlike the case of Kerr BHs where the corrections are from the NLO terms of $M\mu$~\cite{Bao:2022hew,Bao:2023xna}, there are two small parameters $M\mu$ and $a/\ell$ for Kerr-AdS BHs. Below we calculate the NLO contributions arising from both of them.
In Eq.~\eqref{eq:near}, only the LO terms $l(l+1)$ in the coefficients of $z$ and $z^2$ are kept. Now we include the higher order terms and reexpress $V(z)$ in Eq.~\eqref{eq:near_v} as follows:
\begin{align}
V(z)=p^2+(p^2-q^2) z-l^\prime(l^\prime+1) z(z+1),
\end{align}
where $p^2$ is defined in Eq.~\eqref{eq:p} and $q^2$ is given as
\begin{align}
q^2=&\frac{2\omega(r_+ +r_-) \left(a^2+r_+^2\right) (\omega-m \Omega_+)}{r_+-r_-}\nonumber\\
      &+2\mu ^2 r_+^2-4 r_+^2\omega^2-2 \mu ^2 r_- r_+ -p^2.
\end{align}
The $l^\prime$ is defined by the coefficient of $z^2$ in Eq.~\eqref{eq:near_v},
\begin{align}
l^\prime(l^\prime+1)=&\Lambda_{lm}+3\mu^2r_+^2-2\mu^2r_+r_- -4\omega^2r_+^2\nonumber\\
      &-2\omega\left(r_+^2+a^2\right)\left(\omega-m\Omega_+\right). \label{eq:lprime}
\end{align}
It can still be written as
\begin{align}
l^\prime=l+\epsilon,
\end{align}
where the small correction $\epsilon$ is expressed as
\begin{align}\label{eq:epsilon}
     \epsilon=&\frac{1}{2l+1}\left[\Lambda_{lm}-l\left(l+1\right)+3\mu^2r_+^2-2\mu^2r_+r_-\right.\nonumber\\
      &\left.-2\omega\left(r_+^2+a^2\right)\left(\omega-m\Omega_+\right)-4\omega^2r_+^2\right],
\end{align}
which includes the terms at the order of $\mathcal{O}\left(1/\ell, \mu^2 \right)$. Different from the LO calculation, the $\epsilon$ is kept in the final result at the NLO.

With this new $V(x)$, the solution of the radial equation near the horizon is
\begin{align}
	R(r)\propto &\frac{\left(r-r_{-}\right)^{iq}}{\left(r-r_{+}\right)^{i p}}{}_{2}F_{1}\bigg( -l'-i p+iq,\nonumber\\
	& l'+1-i p+iq;1-2ip;-\frac{r-r_+}{r_+-r_-}\bigg).
\end{align}
The large-$r$ behavior of this solution is
\begin{align}
	R\propto &\frac{(r_+-r_-)^{-l^{\prime}}\Gamma\left( 2l^{\prime}+1\right) }{\Gamma\left( l^{\prime}+1-i p-iq\right) \Gamma\left( l^{\prime}+1-i p+iq\right) }r^{l'} \nonumber\\
	&+\frac{\left(r_+-r_-\right) ^{l'+1}\Gamma\left( -2l^{\prime}-1\right) }{\Gamma\left( -l^{\prime}-i p-iq\right) \Gamma\left( -l^{\prime}-i p+iq\right) }r^{-l^{\prime}-1} .
\end{align}
This asymptotic behavior is independent on the sign of $q$. 

Next, we turn to the region far away from the horizon. 
The LO equation in Eq.~\eqref{eq:farLO} does not depend on any BH parameter, implying the solution is insensitive to the BH. By including the NLO in $1/\ell$ expansion, the radial equation far away from the horizon is
\begin{align}\label{eq:far_eq_nlo}
		&\left(1+\frac{r^2}{\ell^2}\right)\partial_{r}^{2}R\left(r\right)+{{2r\left(\frac{2}{\ell^{2}}+\frac{1}{r^2}\right)\partial_{r}R(r)}}\nonumber\\
		&+\left(\frac{\omega^{2}+2am\omega/\ell^2}{1+r^2/\ell^2} -\frac{l^\prime(l^\prime+1)}{r^2}-\mu^{2}\right)R\left(r\right)=0,
\end{align}
in which the NLO term depends on the BH spin $a$. The equation and solution are of the same form with Eq.~\eqref{eq:farLO} and Eq.~\eqref{eq:far_LO} except that the $l$ is replaced by $l^\prime$ and $\omega$ is replaced by $\omega+am/\ell^2$. 
Thus it is straightforward to obtain,
\begin{align}
\frac{l^\prime}{2}+\frac{3}{4}+\frac{\sqrt{9+4\mu^2\ell^2}}{4}+\frac{\omega\ell+am/\ell}{2}=-n-\delta.\label{eq:omega_nlo}
\end{align}

Following the same matching steps, one could obtain $\delta$ with nonzero $\epsilon$,
\begin{widetext}
\begin{equation}
\begin{aligned}
    \delta=&\frac{\left(-1\right) ^{n+1}}{\ell^{2l'+1}}\frac{\left(r_+-r_-\right)^{2l'+1} \Gamma \left(-l'-1/2\right) \Gamma\left( -2l'-1\right) \Gamma\left( n+l'+3/2+\sqrt{9+4\mu^2\ell^2}/2\right)  }{n!\Gamma\left( -l'-1/2-n\right) \Gamma \left(l'+1/2\right) \Gamma \left( 2l'+1\right) \Gamma\left( n+1+\sqrt{9+4\mu^2\ell^2}/2\right)} \\
	&\times\frac{\Gamma \left(l'-i p-iq+1\right) \Gamma \left(l'-i p+iq+1\right)}{\Gamma \left(-l'-i p-iq\right) \Gamma \left(-l'-i p+iq\right)}. 
\end{aligned}
\end{equation}
\end{widetext}
We treat both $\delta$ and $\epsilon$ as perturbation, which leads to
\begin{align}
\omega_0&=\frac{4n+2l+3+\sqrt{9+4\mu^2\ell^2}}{2\ell},\\
\omega_\text{NLO}&=\omega_0+\frac{\epsilon(\omega_0)}{\ell}-\frac{am}{\ell^2}+\frac{2}{\ell}\delta(\omega_0).\label{eq:nlors}
\end{align}
The second term on the right side of Eq.~\eqref{eq:nlors}  comes from the correction to $\ell$ as shown in Eq.~\eqref{eq:epsilon},  which is at the order of $\mathcal{O}(1/\ell^2,\mu^2/\ell)$, while the third term is from the NLO term included in Eq.~\eqref{eq:far_eq_nlo}, which is at the order of $\mathcal{O}(1/\ell^2)$. Compared to the $\omega_0$, which is at the order of $\mathcal{O}(1/\ell,\mu)$, these two terms are NLO corrections. They are both real and sensitive to the BH mass and spin, which results in that the NLO real part is dependent on the BH parameters compared to $\omega_\text{LO}$.

In the next section, we study this dependence with the numerical result. It is noteworthy that the $\delta(\omega_0)$ is not purely imaginary with $\epsilon$ kept in the expression. However, its contribution to $\mathrm{Re}(\omega)$ is at the order of $\epsilon/\ell^{2l+2}$ which can be neglected.


Figure.~\ref{fig:lo_nlo} presents the LO and NLO of the $\mathrm{Im}(\omega)$ and their relative differences for different $\ell$. Both LO and NLO curves have the same qualitative behavior. They increase with $\ell^{-1}$ and drop rapidly to below zero after reaching a maximum. The critical $\ell$ where the curve crosses the horizontal axis is determined by the superradiance condition in Eq.~\eqref{eq:superradiance}. The difference between the LO and the NLO results is generally larger for smaller $\ell$ and larger $a$. For $a=0.99$, the maximum difference exceeds 100\%, indicating the importance of the NLO corrections when $a$ is particularly large.

\section{Numerical results}\label{sec:num}

\begin{table*}
\caption{\label{tab1} The eigenfrequencies obtained with the LO and NLO approximations, the numerical method and the result from Ref.~\cite{Uchikata:2009zz}. The parameters are fixed at $n=0$ with $\ell=1$ and $r_+=0.1$.}
\begin{ruledtabular}
    \begin{tabular}{cccccc}
$(l,m)$& $a$&Ref.~\cite{Uchikata:2009zz}&Numerical&NLO&LO \\
		\hline
  $(0,0)$&$0.01$&$2.69-1.03\times10^{-1} i $&$2.69- 1.02\times10^{-1} i$&$2.46-6.32\times10^{-3} i$&$ 3-5.14\times10^{-2} i$ \\
&$0.05$&$2.60-1.64\times10^{-1} i$&$2.60-1.60\times10^{-1} i$&$2.43-4.76\times10^{-2} i$&$3-6.37\times10^{-2} i$ \\
&$0.10$&$2.43-4.27\times10^{-1} i$&$2.47-1.48\times10^{-1} i$&$ 2.34-4.11\times10^{-1} i$&$3-1.02\times10^{-1} i$\\
\hline
  $(1,1)$&$0.01$&$3.79-1.28\times10^{-3} i$&$3.79- 1.26\times10^{-3} i$&$3.68-1.12\times10^{-3} i$&$ 4-3.48\times10^{-4} i$ \\
&$0.05$&$3.72+7.67\times10^{-5} i$&$3.72+7.90\times10^{-5} i$&$3.63+7.85\times10^{-5} i$&$4-6.02\times10^{-7} i$ \\
&$0.10$&$3.53+3.11\times10^{-4} i$&$3.53+3.19\times10^{-4} i$&$ 3.54+1.17\times10^{-4} i$&$4+2.35\times10^{-5} i$\\
\hline
  $(1,0)$&$0.01$&$3.80-2.21\times10^{-3} i$&$3.80- 2.27\times10^{-3} i$&$3.69-2.04\times10^{-3} i$&$ 4-5.60\times10^{-4} i$ \\
&$0.05$&$3.73-4.47\times10^{-3} i$&$3.73-4.31\times10^{-3} i$&$3.67-4.09\times10^{-3} i$&$4-6.66\times10^{-4} i$ \\
&$0.10$&$3.49-6.12\times10^{-2} i$&$3.50-6.05\times10^{-2} i$&$ 3.70-7.02\times10^{-2} i$&$4-1.74\times10^{-3} i$\\
\hline
 $(1,-1)$& $0.01$&$3.80-3.53\times10^{-3} i$&$3.80- 3.59\times10^{-3} i$&$3.70-3.46\times10^{-3} i$&$ 4-8.54\times10^{-4} i$ \\
&$0.05$&$3.74-2.73\times10^{-2} i$&$3.74-2.70\times10^{-2} i$&$3.73-3.51\times10^{-2} i$&$4-3.86\times10^{-3} i$ \\
&$0.10$&$3.53-2.14\times10^{-1} i$&$3.58-1.49\times10^{-1} i$&$ 3.65-5.93\times10^{-1} i$&$4-1.95\times10^{-2} i$
	\end{tabular}
 \end{ruledtabular}
 \end{table*}

 \begin{figure}[!tb]
	\centering
	\includegraphics[width=0.4\textwidth]{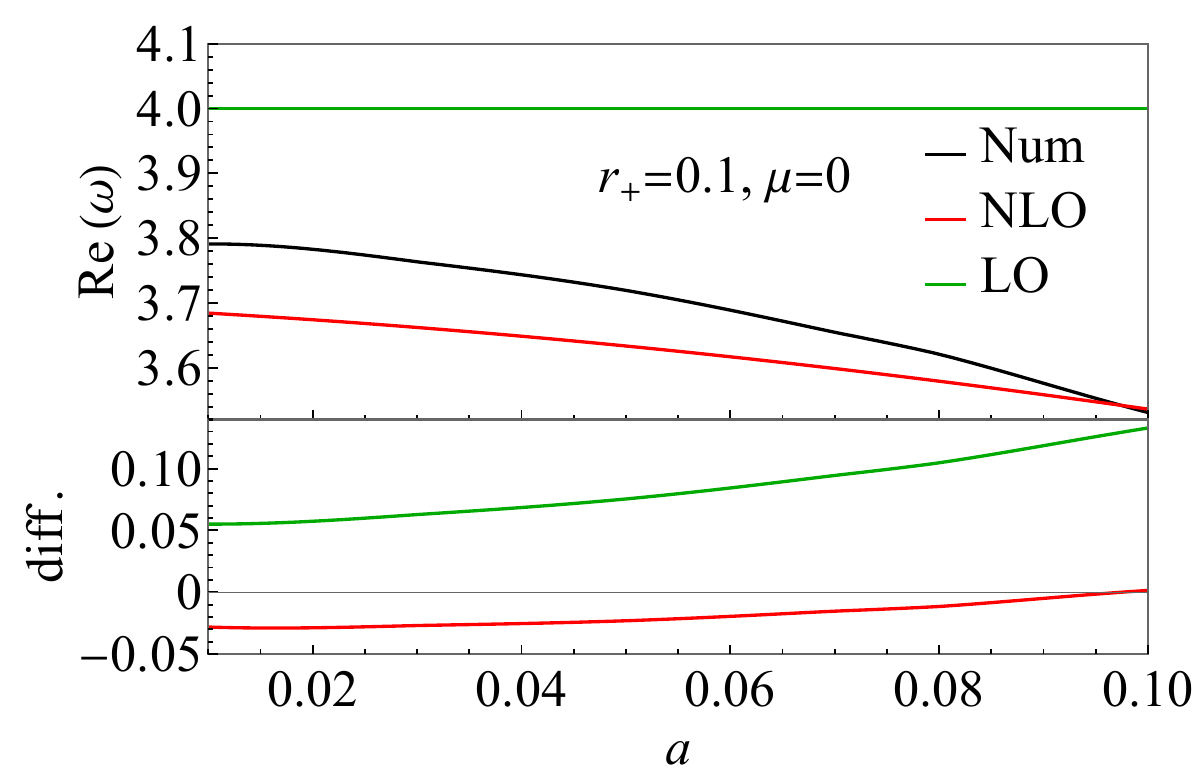}\\
	\includegraphics[width=0.4\textwidth]{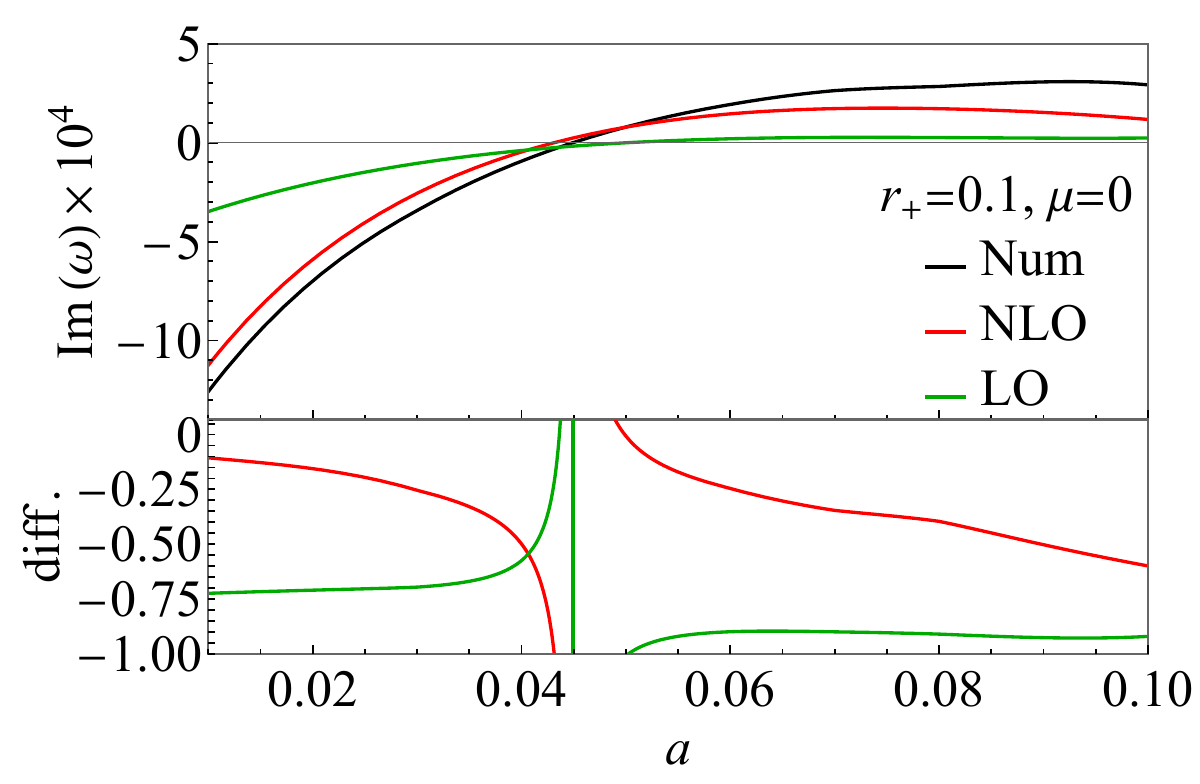}
	\caption{The analytical and numerical results as well as their relative difference of the eigenfrequency $\omega$ as functions of BH spin $a$ with $r_+=0.1$, $\mu=0$, $n=0$, and $l=m=1$. The normalization is chosen as $\ell=1$. The bottom panel of each figure presents the relative difference defined as $\mathrm{Re}(\omega_\mathrm{ana})/\mathrm{Re}(\omega_\mathrm{num})-1$ or $\mathrm{Im}(\omega_\mathrm{ana})/\mathrm{Im}(\omega_\mathrm{num})-1$.\label{fig:num_nlo_lo_mu0}}
\end{figure}
\begin{figure}[!tb]
	\centering
	\includegraphics[width=0.4\textwidth]{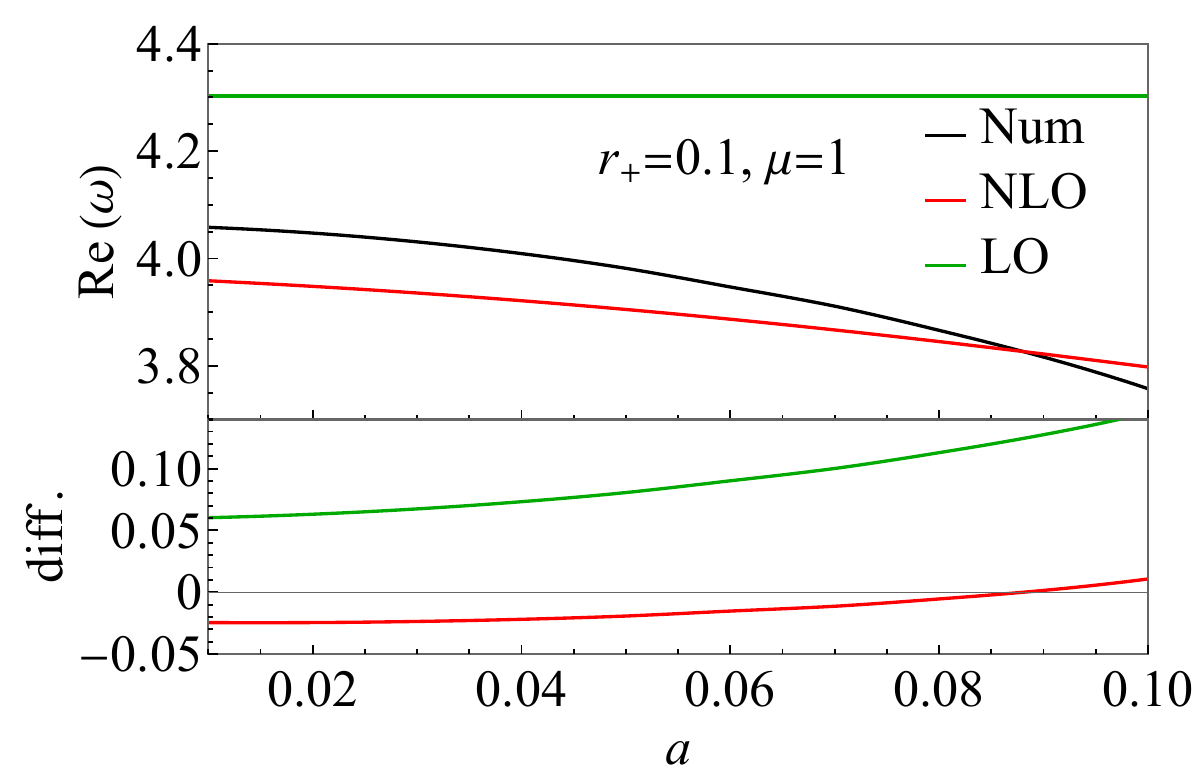}
	\includegraphics[width=0.4\textwidth]{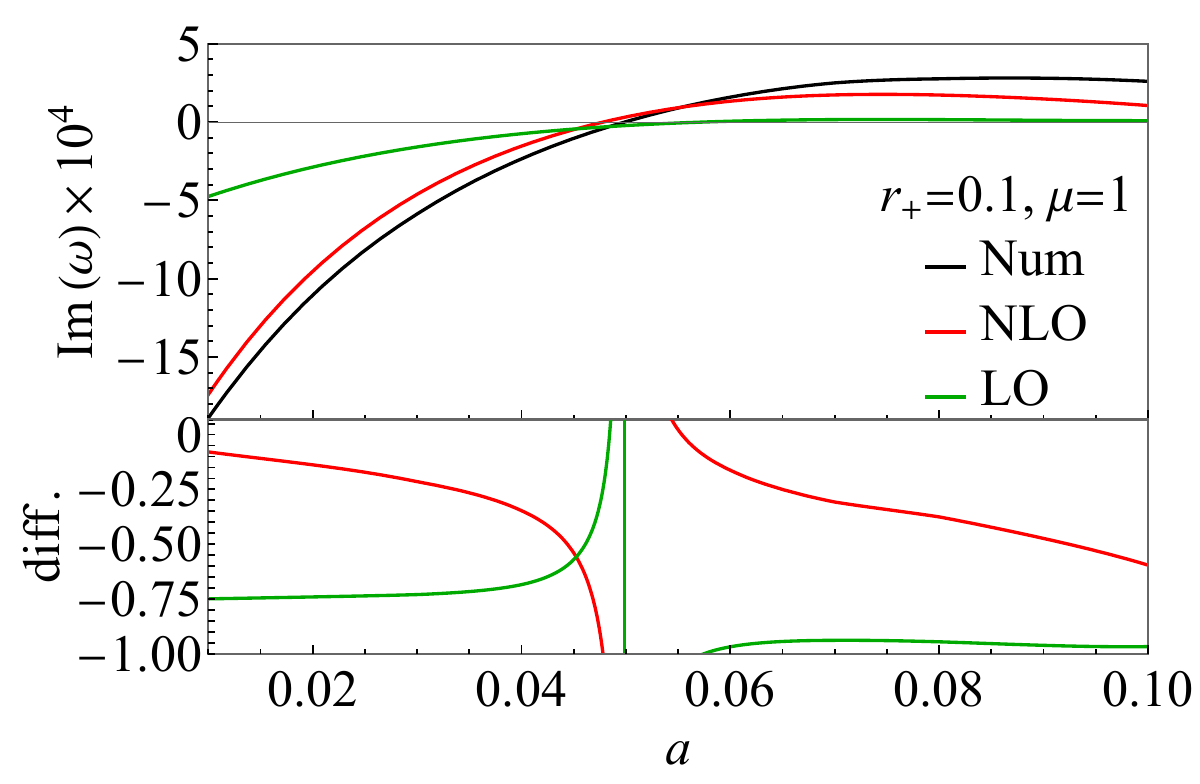}
	\caption{The analytical and numerical results as well as their relative differences of the eigenfrequency $\omega$ as functions of BH spin $a$ with $r_+=0.1$, $\mu=1$, $n=0$, and $l=m=1$. The normalization is chosen as $\ell=1$. The bottom panel of each figure presents the relative difference defined as $\mathrm{Re}(\omega_\mathrm{ana})/\mathrm{Re}(\omega_\mathrm{num})-1$ or $\mathrm{Im}(\omega_\mathrm{ana})/\mathrm{Im}(\omega_\mathrm{num})-1$.\label{fig:num_nlo_lo_mu1} }
\end{figure}

In this section, we calculate the eigenfrequency numerically, which is then compared to the analytic LO and NLO results above. The result for the $\mu=0$ case has been calculated in Ref.~\cite{Uchikata:2009zz}, which clearly shows the dependence of $\text{Re}(\omega)$ on the BH spin.  Below we solve the case of $\mu\ne 0$ with the shooting method.

As discussed in Sec.~\ref{sec:lo}, the radial function $R(r)$ of a bound state must satisfy boundary conditions,
\begin{align}
	\lim_{r\to r_+} R(r)\propto\left( \frac{r-r_+}{r-r_-}\right) ^{-ip}.\label{eq:boundary} 
\end{align}
in the neighborhood of $r_+$ and 
\begin{align}\label{eq:boundary_infty}
\lim_{r\to\infty}R(r)\propto r^{-3}.
\end{align}
at large $r$. We employ the shooting method to solve this eigenvalue problem. Starting with Eq.~\eqref{eq:boundary} near $r_+$, one could evaluate $R(r)$ numerically to the large $r$ region with Eq.~\eqref{eq:radial}. The $R(r)$ will satisfy Eq.~\eqref{eq:boundary_infty} if and only if $\omega$ is the eigenfrequency. The $\Lambda_{lm}$ in Eq.~\eqref{eq:radial} should be solved from the angular equation with continued fraction method~\cite{Leaver:1985ax,Uchikata:2009zz}. In this work, Eq.~\eqref{eq:lambdalm} is adopted for efficiency. We have verified numerically that this approximation works well for all $l\geq1$. 

In Table~\ref{tab1}, we present the results obtained with both LO and NLO approximations, as well as the  numerical method for the case when $\mu=0$ and $n=0$. To compare with the numerical results in Ref.~\cite{Uchikata:2009zz}, we take $\ell=1$ as the normalization in this section.  Our numerical results agree very well with those from Ref.~\cite{Uchikata:2009zz} for most parameter sets, especially for the real part of the eigenfrequencies. However, there is a significant discrepancy in the imaginary part for extremely spinning BHs characterized by $a\sim r_+$ when $l=m=0$. We do not have a definitive explanation for this discrepancy. One possible reason could be that Eq.~\eqref{eq:lambdalm} is not a good approximation for $l=m=0$. 

Moreover, Table~\ref{tab1} exhibits a significant difference between the LO approximation and the numerical result. In particular, the real part of the LO results in Eq.~\eqref{eq:omega_lo} is independent of the BH spin. Comparably, the NLO results have a remarkable improvement. The $\text{Re}\left(\omega_\text{NLO}\right)$ is closely aligned with the numerical result, and the discrepancies of the imaginary parts are also reduced for $l\geq 1$. 

The disagreement in the $l=m=0$ case is understandable. In both the LO and NLO approximations, we have assumed $\Lambda_{lm}\sim l(l+1)\gg r_+^2/\ell^2$, which is invalid for the case  $l=m=0$. In addition, for the extremely spinning BHs $a\sim r_+$, some of the neglected terms are enhanced by $1/(r_+-r_-)$ in Eq.~\eqref{eq:near_v}. It explains the larger discrepancy with a larger value of $a$.

In Fig.~\ref{fig:num_nlo_lo_mu0}, we plot the LO, NLO, and numerical results for both the real part (top panel) and the imaginary part (bottom panel) of the eigenfrequency, as well as the relative differences for $l=m=1$. The $\mathrm{Im}(\omega)$ changes sign at $a_c\simeq 0.046$, which agrees with the critical value in Eq.~\eqref{eq:a_critical} for $\mu=0$. Except in the neighborhood near the critical value, the differences between NLO and numerical results are within 5\% for the real part and  60\% for the imaginary part, which are significantly smaller than the discrepancies between the LO and numerical results. 

The results with $\mu=1$ are presented in Fig.~\ref{fig:num_nlo_lo_mu1}. The NLO real part as a function of $a$ is consistent with the numerical result, with a relative difference within 5\%. On the other hand, the relative difference between the NLO imagery part and the numerical result is less than 60\% in most regions. The imaginary part changes sign at $a\sim 0.051$, consistent with Eq.\eqref{eq:a_critical}.

\section{Summary}\label{sec:sum}

In this work, we have investigated the scalar superradiant instability of Kerr-AdS BHs. We reproduce the LO result and extend the analysis to the NLO. We further calculate the eigenfrequency numerically with the shooting method to check the reliability of the NLO results.

Employing matched asymptotic expansion techniques, we reproduced the LO result for Kerr-AdS BHs. Our result is consistent with that in Ref.~\cite{Delice:2015zga} for $\mu\neq0$. Nonetheless, the obtained imaginary part of the eigenfrequency has an extra overall factor of 1/2 compared to the result in Ref.~\cite{Cardoso:2004hs} for the case of $\mu=0$. This factor is from the regulation of the ill-defined Gamma functions. Particularly, the real part is independent of the BH mass and spin, which is a significant concern of the LO result. We found this problem is due to the utilization of the free AdS solution in the region far away from the BH. To improve the result, we introduced the $\mathcal{O}\left((M\mu)^2\right)$ and $\mathcal{O}\left(1/\ell^2\right)$ contributions in our analysis and derived the NLO contribution in a compact form. Notably, the asymptotic equation in the large-$r$ region is sensitive to the BH spin at NLO. The real part of the eigenfrequency then relies on the BH spin in this order. 

We further performed numerical calculations with self-written code to study the accuracy of the NLO approximation. Our numerical results are consistent with those in Ref.~\cite{Uchikata:2009zz} for the case of $\mu=0$. Compared to the numerical results, the NLO expression presents a much better convergence.  Specifically, the percentage error of the $\text{Re}(\omega)$ is within 5\%, while the discrepancy of the imaginary part is within 50\% in most parameter space except when the BH spin is close to the critical $a\to a_c$. Therefore, we conclude the obtained NLO result provides a more accurate description of the scalar superradiant instability of Kerr-AdS BHs.

\begin{acknowledgments}
This work is supported in part by the National Natural Science Foundation of China (NSFC) under Grants No. 12447105 and No. 12075136 and the Natural Science Foundation of Shandong Province under Grant No. ZR2020MA094.
\end{acknowledgments}




\begin{thebibliography}{99}
\bibitem{Penrose:1969pc}
R.~Penrose,
Riv. Nuovo Cimento. \textbf{1}, 252 (1969).

\bibitem{Huang:2018qdl}
Y.~Huang, D.~J.~Liu, X.~h.~Zhai and X.~z.~Li,
Phys. Rev. D \textbf{98}, 025021 (2018).

\bibitem{Boskovic:2018lkj}
M.~Boskovic, R.~Brito, V.~Cardoso, T.~Ikeda, and H.~Witek,
Phys. Rev. D \textbf{99}, 035006 (2019).

\bibitem{Franzin:2021kvj}
E.~Franzin, S.~Liberati , and M.~Oi,
Phys. Rev. D \textbf{103}, 104034 (2021).

\bibitem{Garcia-Saenz:2021uyv}
S.~Garcia-Saenz, A.~Held , and J.~Zhang,
Phys. Rev. Lett. \textbf{127}, 131104 (2021)

\bibitem{Lin:2021ssw}
J.~M.~Lin, M.~J.~Luo, Z.~H.~Zheng, L.~Yin , and J.~H.~Huang,
Phys. Lett. B \textbf{819}, 136392 (2021).

\bibitem{Guo:2021xao}
R.~Z.~Guo, C.~Yuan , and Q.~G.~Huang,
Phys. Rev. D \textbf{105}, 064029 (2022).

\bibitem{Arvanitaki:2010sy}
A.~Arvanitaki , and S.~Dubovsky,
Phys. Rev. D \textbf{83}, 044026 (2011).

\bibitem{Herdeiro:2021znw}
C.~A.~R.~Herdeiro, E.~Radu , and N.~M.~Santos,
Phys. Lett. B \textbf{824}, 136835 (2022).

\bibitem{Brito:2015oca}
R.~Brito, V.~Cardoso , and P.~Pani,
Physics,''
Lect. Notes Phys. \textbf{906}, 1 (2015).
2020,
ISBN 978-3-319-18999-4, 978-3-319-19000-6, 978-3-030-46621-3, 978-3-030-46622-0

\bibitem{Zel'dovich:1971}
Ya. B. Zel'dovich, JETP Lett. {\bf 14}, 180 (1971).

\bibitem{Bardeen:1972fi}
J.~M.~Bardeen, W.~H.~Press , and S.~A.~Teukolsky,
Astrophys. J. \textbf{178}, 347 (1972).

\bibitem{Starobinsky:1973aij}
A.~A.~Starobinsky,
Sov. Phys. JETP \textbf{37}, 28 (1973).

\bibitem{Press:1972zz}
W.~H.~Press , and S.~A.~Teukolsky,
Nature(London) \textbf{238}, 211 (1972).

\bibitem{Cardoso:2004nk}
V.~Cardoso, O.~J.~C.~Dias, J.~P.~S.~Lemos , and S.~Yoshida,
Phys. Rev. D \textbf{70}, 044039(E) (2004)
[erratum: Phys. Rev. D \textbf{70}, 049903 (2004)].

\bibitem{Strafuss:2004qc}
M.~J.~Strafuss , and G.~Khanna,
Phys. Rev. D \textbf{71}, 024034 (2005).

\bibitem{Hod:2009cp}
S.~Hod , and O.~Hod,
Phys. Rev. D \textbf{81}, 061502 (2010).

\bibitem{Rosa:2009ei}
J.~G.~Rosa,
J. High Energy Phys. \textbf{06} (2010) 015.

\bibitem{Witek:2010qc}
H.~Witek, V.~Cardoso, C.~Herdeiro, A.~Nerozzi, U.~Sperhake , and M.~Zilhao,
Phys. Rev. D \textbf{82}, 104037 (2010).

\bibitem{Dolan:2012yt}
S.~R.~Dolan,
Phys. Rev. D \textbf{87}, 124026 (2013).

\bibitem{Herdeiro:2013pia}
C.~A.~R.~Herdeiro, J.~C.~Degollado , and H.~F.~R\'unarsson,
Phys. Rev. D \textbf{88}, 063003 (2013).

\bibitem{Lee:2011ez}
J.~P.~Lee,
J. High Energy Phys. \textbf{01} (2012) 091.

\bibitem{Aliev:2014aba}
A.~N.~Aliev,
J. Cosmol. Astropart. Phys. \textbf{11}, 029 (2014).

\bibitem{Degollado:2013bha}
J.~C.~Degollado , and C.~A.~R.~Herdeiro,
Phys. Rev. D \textbf{89}, 063005 (2014).

\bibitem{Hod:2013fvl}
S.~Hod,
Phys. Rev. D \textbf{88}, 064055 (2013).

\bibitem{Li:2014gfg}
R.~Li, J.~K.~Zhao , and Y.~M.~Zhang,
Commun. Theor. Phys. \textbf{63},  569 (2015).

\bibitem{Li:2014fna}
R.~Li , and J.~Zhao,
Phys. Lett. B \textbf{740}, 317 (2015).

\bibitem{Cardoso:2006wa}
V.~Cardoso, O.~J.~C.~Dias , and S.~Yoshida,
Phys. Rev. D \textbf{74}, 044008 (2006).

\bibitem{Kunduri:2006qa}
H.~K.~Kunduri, J.~Lucietti , and H.~S.~Reall,
Phys. Rev. D \textbf{74}, 084021 (2006).

\bibitem{Kodama:2009rq}
H.~Kodama, R.~A.~Konoplya , and A.~Zhidenko,
Phys. Rev. D \textbf{79}, 044003 (2009).

\bibitem{Kodama:2007sf}
H.~Kodama,
Prog. Theor. Phys. Suppl. \textbf{172}, 11 (2008).

\bibitem{Cardoso:2013pza}
V.~Cardoso, \'O.~J.~C.~Dias, G.~S.~Hartnett, L.~Lehner , and J.~E.~Santos,
J. High Energy Phys. \textbf{04}, (2014) 183.

\bibitem{Murata:2008xr}
K.~Murata,
Prog. Theor. Phys. \textbf{121}, 1099 (2009).

\bibitem{Aliev:2008yk}
A.~N.~Aliev , and O.~Delice,
Phys. Rev. D \textbf{79}, 024013 (2009).

\bibitem{Wang:2014eha}
M.~Wang , and C.~Herdeiro,
Phys. Rev. D \textbf{89},  084062 (2014).

\bibitem{Delice:2015zga}
\"O.~Delice , and T.~Dur\u{g}ut,
Phys. Rev. D \textbf{92}, 024053 (2015).

\bibitem{Aliev:2015wla}
A.~N.~Aliev,
Eur. Phys. J. C \textbf{76}, 58 (2016).

\bibitem{Hawking:1999dp}
S.~W.~Hawking , and H.~S.~Reall,
Phys. Rev. D \textbf{61}, 024014 (2000).

\bibitem{Cardoso:2004hs}
V.~Cardoso , and O.~J.~C.~Dias,
Phys. Rev. D \textbf{70}, 084011 (2004).

\bibitem{Uchikata:2009zz}
N.~Uchikata, S.~Yoshida , and T.~Futamase,
Phys. Rev. D \textbf{80}, 084020 (2009).

\bibitem{Gubser:2008px}
S.~S.~Gubser,
Phys. Rev. D \textbf{78}, 065034 (2008).

\bibitem{Detweiler:1980uk}
S.~L.~Detweiler,
Phys. Rev. D \textbf{22}, 2323 (1980).

\bibitem{Li:2012rx}
R.~Li,
Phys. Lett. B \textbf{714}, 337 (2012).

\bibitem{Carter:1968rr}
B.~Carter,
Phys. Rev. \textbf{174}, 1559 (1968).

\bibitem{Carter:1968ks}
B.~Carter,
Commun. Math. Phys. \textbf{10}, 280 (1968).

\bibitem{Hawking:1998kw}
S.~W.~Hawking, C.~J.~Hunter , and M.~Taylor,
Phys. Rev. D \textbf{59}, 064005 (1999).

\bibitem{Gibbons:2004ai}
G.~W.~Gibbons, M.~J.~Perry , and C.~N.~Pope,
Classical Quantum Gravity. \textbf{22}, 1503 (2005).

\bibitem{Teukolsky:1972my}
S.~A.~Teukolsky,
Phys. Rev. Lett. \textbf{29}, 1114 (1972).

\bibitem{Suzuki:1998vy}
H.~Suzuki, E.~Takasugi , and H.~Umetsu,
Prog. Theor. Phys. \textbf{100}, 491 (1998).

\bibitem{Cho:2009wf}
H.~T.~Cho, A.~S.~Cornell, J.~Doukas , and W.~Naylor,
Phys. Rev. D \textbf{80}, 064022 (2009).

\bibitem{Berti:2005gp}
E.~Berti, V.~Cardoso , and M.~Casals,
Phys. Rev. D \textbf{73}, 024013 (2006)
[erratum: Phys. Rev. D \textbf{73}, 109902(E) (2006)].

\bibitem{Seidel:1988ue}
E.~Seidel,
Classical Quantum Gravity. \textbf{6}, 1057 (1989).

\bibitem{Ishibashi:2004wx}
A.~Ishibashi , and R.~M.~Wald,
Classical Quantum Gravity. \textbf{21}, 2981 (2004).

\bibitem{Breitenlohner:1982bm}
P.~Breitenlohner , and D.~Z.~Freedman,
Phys. Lett. \textbf{115}B, 197 (1982).

\bibitem{Bao:2022hew}
S.~Bao, Q.~Xu , and H.~Zhang,
Phys. Rev. D \textbf{106}, 064016 (2022).

\bibitem{Bao:2023xna}
S.~S.~Bao, Q.~X.~Xu , and H.~Zhang,
Phys. Rev. D \textbf{107}, 064037 (2023).

\bibitem{Guo:2022mpr}
Y.~d.~Guo, S.~s.~Bao , and H.~Zhang,
Phys. Rev. D \textbf{107}, 075009 (2023).

\bibitem{Leaver:1985ax}
E.~W.~Leaver,
Proc. R. Soc. A \textbf{402}, 285 (1985).
\end{thebibliography}
\end{document}